# Thickness dependence of the properties of epitaxial MgB$_2$ thin films grown by hybrid physical-chemical vapor deposition

A. V. Pogrebnyakov[a]

*Department of Physics and Materials Research Institute,*

*The Pennsylvania State University, University Park, PA 16802*

J. M. Redwing

*Department of Materials Science and Engineering and Materials Research Institute,*

*The Pennsylvania State University, University Park, PA 16802*

J. E. Jones, X. X. Xi, S. Y. Xu, and Qi Li

*Department of Physics and Materials Research Institute,*

*The Pennsylvania State University, University Park, PA 16802*

V. Vaithyanathan and D. G. Schlom

*Department of Materials Science and Engineering and Materials Research Institute,*

*The Pennsylvania State University, University Park, PA 16802*

## Abstract

We have studied the effect of deposition rate and layer thickness on the properties of epitaxial MgB$_2$ thin films grown by hybrid physical-chemical vapor deposition on 4H-SiC substrates. The MgB$_2$ film deposition rate depends linearly on the concentration of B$_2$H$_6$ in the inlet gas mixture. We found that the superconducting and normal-state properties of the MgB$_2$ films are determined by the film thickness, not by the deposition rate. When the film thickness was increased, the transition temperature, $T_c$, increased and the residual resistivity, $\rho_0$, decreased. Above 300 nm, a $T_c$ of 41.8 K, a $\rho_0$ of 0.28 $\mu\Omega$·cm, and a residual resistance ratio $RRR$ of over 30 were obtained. These values represent the best MgB$_2$ properties reported thus far.

For both fundamental studies and electronic applications of the magnesium diboride superconductor [1], it is desirable to have high quality single-crystalline thin films with intrinsic superconducting and normal-state properties. Since its discovery, the best values for the $MgB_2$ properties reported in the literature, in either bulk samples, single crystals, or thin films, are a critical temperature, $T_c$, of 40 K [2], a low residual resistivity, $\rho_0$, of 0.28 $\mu\Omega$·cm [3], and a high residual resistance ratio, *RRR*, of 25 [4]. Studies show that high-*RRR* and low-$\rho_0$ samples have to be prepared using high purity boron and magnesium [5] and proper heat treatment [3]. On the other hand, the $T_c$ near 39 K is relatively insensitive to the normal-state resistivity, likely due to the poor connectivity between grains in high-$\rho_0$ samples [6]. Recently, we have fabricated epitaxial $MgB_2$ thin films with bulk-like properties using hybrid physical-chemical vapor deposition (HPCVD) [7]. The quality of the HPCVD deposited films is reproducibly very high, which allows us to systematically explore the critical materials parameters for $MgB_2$ films. In this letter, we present the results for a series of $MgB_2$ films with different thicknesses and show that the film thickness has a strong influence on the superconducting and normal-state properties. In $MgB_2$ films that are above approximately 3000 Å, the $T_c$, $\rho_0$, and *RRR* values are equal to or exceed the best reported values.

Detailed descriptions of the epitaxial growth of $MgB_2$ by HPCVD are contained in our previous publications [7,8]. Briefly, the HPCVD system consists of a vertical quartz reactor with an inductively heated susceptor. A single crystal substrate is placed on the top surface of the susceptor with magnesium slugs nearby. When the susceptor is heated in a hydrogen flow (400-1000 sccm at a pressure of 100-700 Torr) to 720-760 ºC, a high Mg vapor pressure necessary for the $MgB_2$ growth [9] is generated near the substrate. Since the sticking coefficient of Mg is very low above 300 ºC [10], no Mg film is formed on the substrate. The $MgB_2$ film growth begins when the boron precursor gas, 1000 ppm diborane ($B_2H_6$) in $H_2$, starts to flow into the reactor. In our previous studies, the flow rate of the $B_2H_6$ gas mixture was 25-50 sccm. In the present work, the total gas flow was kept at 450 sccm flow rate and the reactor pressure was 100 Torr. The $B_2H_6$ gas mixture flow rate was varied between 50-250 sccm, which corresponds to a change in the mole fraction of $B_2H_6$ in the inlet gas from $1.1 \times 10^{-4}$ to $5 \times 10^{-4}$. The films were deposited on 4H-SiC substrates at 720 ºC. We have shown previously that SiC is an excellent substrate for growing $MgB_2$ thin films [7]. X-ray diffraction showed that the films are *c*-axis oriented and epitaxial with in-plane alignment between the *a*-axis of $MgB_2$ and the *a*-axis of SiC. The full width at half maximum (FWHM) of the 0002 $MgB_2$ peak was less than 0.30º in $2\theta$ and less than 0.65º in $\omega$ (rocking curve) for the films described. The thickness of the films was measured by a Dektak profilometer over etched edges.

The deposition rate of $MgB_2$ films as a function of the $B_2H_6$ gas mixture flow rate is shown in Fig. 1. The data were obtained from films of different thicknesses that were deposited for different lengths of time. A linear dependence was observed. The deposition rate increases from about 3 Å/s for 50 sccm to about 18 Å/s for 250 sccm of $B_2H_6$ gas mixture flow. This is consistent with the thermodynamic prediction [9] that the $MgB_2$ growth is adsorption-controlled with automatic composition control. As long as the Mg vapor pressure is high enough to keep the $MgB_2$ phase thermodynamically stable and the Mg:B ratio is higher than 1:2, the film composition will automatically be maintained as $MgB_2$. The extra Mg will be in the gas phase and removed by the pump system. The growth rate of $MgB_2$ is then solely determined by the rate at which boron is incorporated into the film. Our result indicates that both the Mg vapor pressure and Mg:B ratio requirements were satisfied in HPCVD for the $B_2H_6$ gas mixture flow as high as 250 sccm.

The $MgB_2$ films deposited with higher $B_2H_6$ gas mixture flow rates generally have larger grain sizes than those grown with lower flow rates. As reported previously [7], films deposited with 25 sccm $B_2H_6$ gas mixture flow are smooth with the root-mean square (RMS) roughness of 25-40 Å, and grain sizes about 0.1-0.2 $\mu$m. $MgB_2$ films with the thickness up to 2000 Å, grown at higher $B_2H_6$ gas mixture flow rates are also smooth but with larger grain sizes. An atomic force microscopy (AFM) image of a 840 Å-thick $MgB_2$ film deposited at 200 sccm of flow rate is presented in Fig. 2(a). It shows that the film is dense and has a relatively flat surface. The RMS roughness is 34 Å, for a 10 x 10 $\mu m^2$ area and 24 Å, for a 2 x 2 $\mu m^2$ area. The average grain size is about 0.5-0.8 $\mu$m. For thicker films the grain size can be larger but the surface becomes rougher. Fig. 2 (b) is an AFM image of a 3400 Å -thick $MgB_2$ film deposited at a flow rate of 250 sccm, which shows a grain as large as 5 $\mu$m in size at the lower left corner of the picture. The RMS roughness of this film is 260 Å in this area.

Resistivity measurements by the four-probe method and ac susceptibility measurements were used to characterize the superconducting and normal-state transport properties of the films. Fig. 3 shows the temperature dependence of the resistivity for a 2250 Å -thick $MgB_2$ film deposited using a 200 sccm flow rate of the $B_2H_6$ gas mixture. The film has a high $T_c$ of 41.7 K with a sharp superconducting transition (< 0.1 K between 90% and 10% of normal-state resistivity) as demonstrated by the inset to Fig. 3. The same narrow transition is also revealed by the ac susceptibility measurement. The residual resistivity of this sample is 0.28 $\mu\Omega$·cm and the *RRR* is ~ 30. To our knowledge, these values represent the best $MgB_2$ properties reported to date.

It is important to determine whether the improved film properties result from the higher deposition rate or the increased film thickness. For this purpose, a series of films were deposited at different deposition rates and for different lengths of time. The thickness dependence of $T_c$ of the $MgB_2$ films made at different $B_2H_6$ gas mixture flow rates is plotted in Fig. 4(a). As can be seen in the figure, there is no dependence of $T_c$ on the flow rate, thus the deposition rate, but a clear dependence on the film thickness. The $T_c$ value increases from 40.3 K to 41.8 K as the thickness increases from 420 Å, to above 3000 Å, an increase of 3.7%. At the same time, the residual resistivity decreases with increasing film thickness as shown in Fig. 4(b). In the thickness range of 420 to above 3000 Å, $\rho_0$ decreases from 1.3 to 0.28 $\mu\Omega\cdot$cm, a drop of 80%. The *RRR* values for the low-$\rho_0$ samples are as high as over 30. Evidently, the critical parameter is the film thickness, not the film deposition rate, in obtaining the high $T_c$ and low $\rho_0$ in the MgB$_2$ films. In both (a) and (b) of Fig. 4, a saturation seems to have been reached above a thickness of about 3000 Å.

The origin of the thickness dependence of the film properties is not clear at present. The low values of the residual resistivity indicate that the films are very pure. This may be attributed to the absence of substantial MgO contamination, due to the reducing hydrogen ambient in the deposition process of HPCVD. Further, the resistivity difference, $\Delta\rho \equiv \rho(300K) - \rho_0$, is small and does not depend on the film thickness, as shown by the inset to Fig. 4(b). As pointed out by Rowell *et al*. [12], $\Delta\rho$ reflects the connectivity between grains. The small value of $\Delta\rho$ and its independence on the film thickness indicate that all the samples studied are fully coalesced and have well connected growth columns. This suggests that the grain size is not likely to be the direct cause of the thickness dependence of $T_c$. Although excess Mg at the grain boundaries may lead to high *RRR* values [13], it cannot explain the high $T_c$ and the thickness dependence, and both thermodynamic theory [9] and experiment [14] show that a Mg-rich MgB$_2$ phase does not exist. A possible explanation of higher $T_c$ is strain in the film. The x-ray diffraction measurement on a 2300 Å-thick film showed a lattice constant of $a = 3.095 \pm 0.015$ Å, which is slightly larger than the value of 3.086 Å, reported for bulk MgB$_2$ [1]. The measured $c$ lattice constant was $3.515 \pm 0.001$ Å, which is slightly smaller than the bulk value of 3.524 Å [1]. This suggests that the films are under tensile in-plane epitaxial strain. Hur *et al*. have reported a higher-than-bulk $T_c$ in MgB$_2$ films on boron crystals and suggested that it is possibly due to tensile strain [2]. Yildirim and Gülseren have predicted an increase in $T_c$ by the *c*-axis compression by first-principle calculations [15]. Further studies are currently under way to better understand the nature of the strain in the MgB$_2$ films and to establish its correlation with film properties.

In conclusion, we found that the deposition rate of MgB$_2$ films by HPCVD is proportional to the B$_2$H$_6$ gas mixture flow rate, consistent with the thermodynamic prediction of the adsorption-controlled growth of MgB$_2$. The high deposition rate leads to larger grain sizes and the film roughness increases with the film thickness. The superconducting and normal-state transport properties, however, do not depend on the deposition rate but rather on the MgB$_2$ film thickness. Larger film thickness results in higher $T_c$, lower $\rho_0$, and higher *RRR*. At above 3000 Å, the best values are $T_c$ = 41.8 K, $\rho_0$ = 0.28 $\mu\Omega\cdot$cm, and *RRR* above 30. Understanding these thickness dependencies may provide insights into ways to further increase $T_c$ in MgB$_2$ or other diboride systems.

This work is supported in part by ONR under grant Nos. N00014-00-1-0294 (Xi) and N0014-01-1-0006 (Redwing), by NSF under grant Nos. DMR-0103354 (Xi) DMR-9876266 and DMR-9972973 (Li), and by DOE through grant DE-FG02-97ER45638 (Schlom).

# References

a) Electronic address: avp11@psu.edu

Figure Captions

Fig. 1.  Deposition rate of MgB$_2$ films by HPCVD as a function of the B$_2$H$_6$ gas mixture flow rate. The data are from films deposited for different lengths of time.

Fig. 2.  Atomic force microscopy images of (a) a 840 Å-thick MgB$_2$ film deposited at 200 sccm of B$_2$H$_6$ gas mixture flow rate, and (b) a 3400 Å-thick MgB$_2$ film deposited at 250 sccm of B$_2$H$_6$ gas mixture flow rate.

Fig. 3.  Resistivity versus temperature for a 2250 Å-thick MgB$_2$ film deposited at 200 sccm of B$_2$H$_6$ gas mixture flow rate. The inset shows details at the superconducting transition.

Fig. 4.  Thickness dependence of (a) $T_c$, and (b) $\rho_0$ of MgB$_2$ films made at different B$_2$H$_6$ gas mixture flow rates. The inset to (b) shows the thickness dependence of $\Delta\rho \equiv \rho(300K) - \rho_0$.

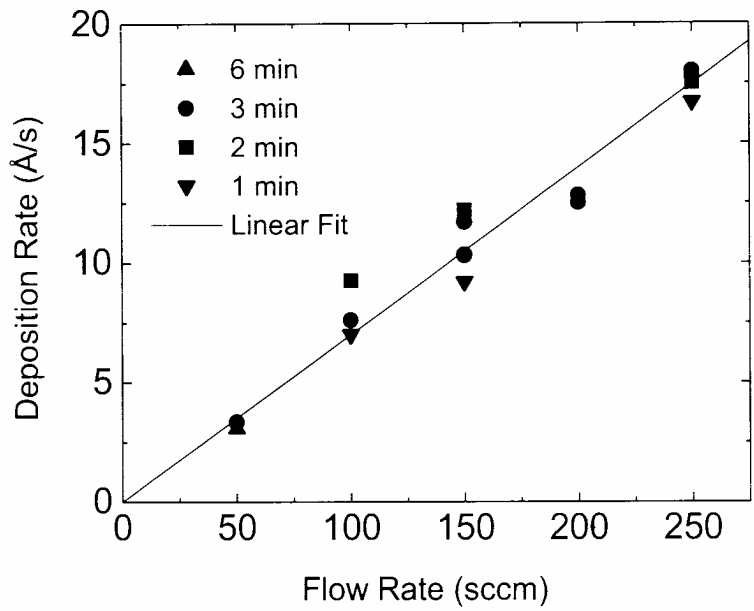

Figure 1 of 4
Pogrebnyakov et al.

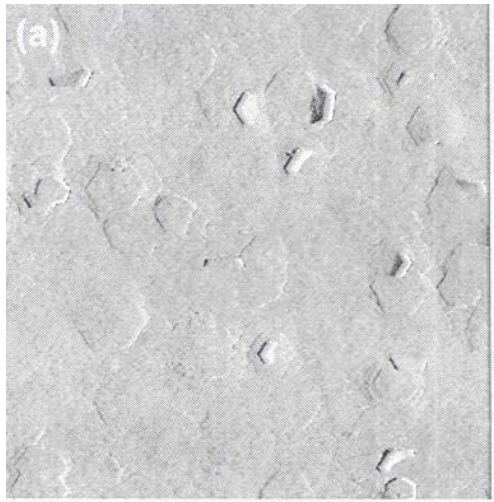 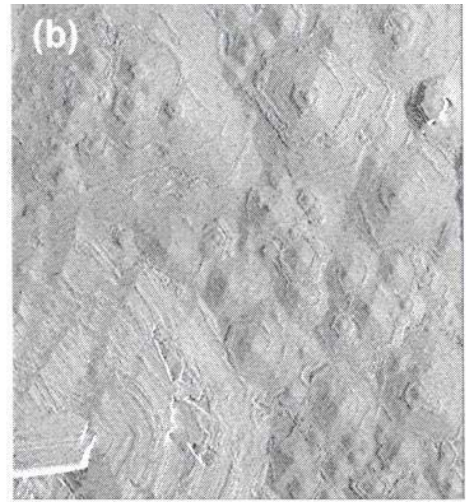

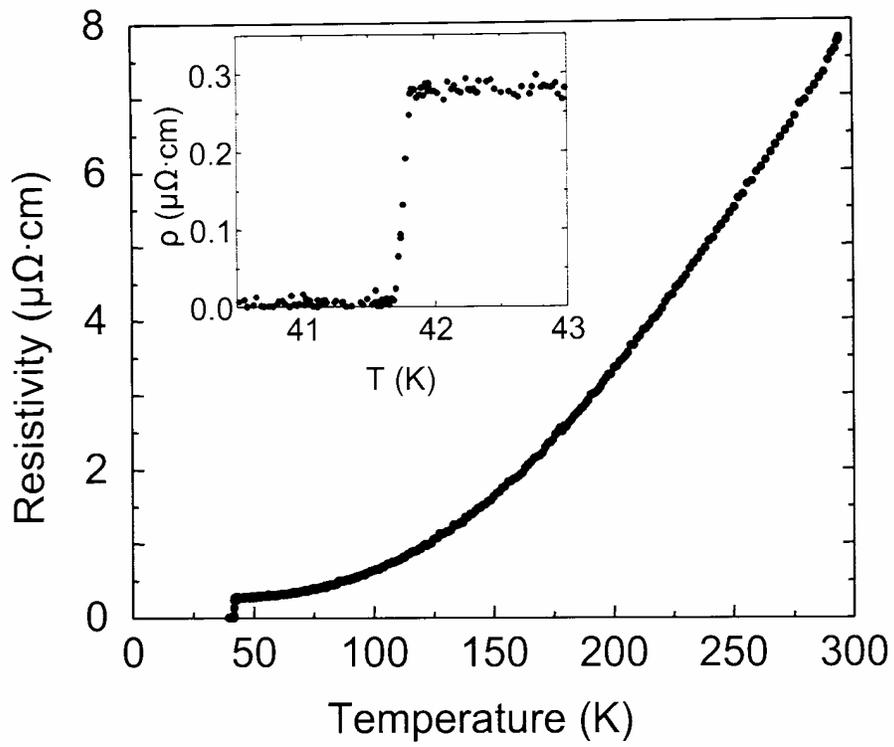

Figure 3 of 4
Pogrebnyakov et al.

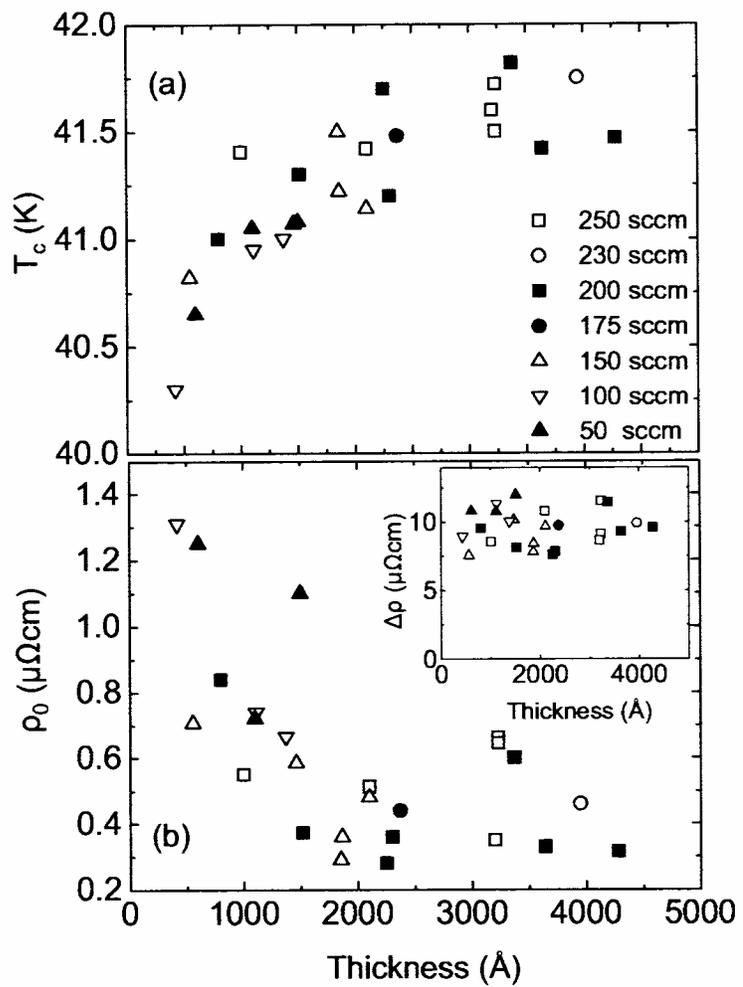

Figure 4 of 4
Pogrebnyakov et al.